\documentclass[iop]{emulateapj}
\usepackage{amsmath}

\slugcomment{Not to appear in Nonlearned J., 45.}

\shorttitle{A New Mechanism for Weakly Magnetized Core-Collapse}
\shortauthors{Sawai \& Yamada}

\begin{document}

\title{Influence of Magnetorotational Instability on Neutrino
  Heating: A New Mechanism for Weakly Magnetized
  Core-Collapse Supernovae}

\author{Hidetomo Sawai\altaffilmark{1} and Shoichi Yamada\altaffilmark{1}} 

\email{hsawai@heap.phys.waseda.ac.jp}

\altaffiltext{1}{Waseda University, Shinjuku, Tokyo 169-8555, Japan}

\begin{abstract}
We investigated the impacts of magnetorotational instability (MRI) on the
dynamics of weakly magnetized, rapidly rotating core-collapse by
conducting high resolution MHD simulations in axisymmetry with
simplified neutrino transfer. We found that an initially
sub-magnetar class magnetic field is drastically amplified by MRI and
substantially affects the dynamics thereafter. Although the magnetic
pressure is not strong enough to eject matter, the amplified magnetic
field efficiently transfers angular momentum from higher to lower
latitudes, which causes the expansion of the heating region at low latitudes
due to the extra centrifugal force. This then enhance the efficiency
of neutrino heating and eventually leads to neutrino-driven explosion.
This is a new scenario of core-collapse supernovae that has never been
demonstrated by numerical simulations so far.
\end{abstract}

\keywords{supernovae: general --- magnetohydrodynamics (MHD) ---
  Instabilities --- methods: numerical  --- stars: magnetars}

\section{Introduction}
At present neutrino heating is considered to be the most promising
candidate for the as-yet-unknown explosion mechanism of core-collapse
supernovae (CCSNe). Recent state-of-the-art simulations with detailed
neutrino transport have not yet succeeded in reproducing
the canonical explosion energy of $10^{51}$~erg, though
\citep[e.g.,][]{suw10,mue12,bru13}. Other effects to enhance or drive
explosion may hence be required. 

Magnetic field and rotation may bring such
effects. Although the magnetic field and rotation of progenitors at the
pre-collapse stage are not well-understood, some of recent theoretical
and observational studies indicate that their strengths may spread
over a wide range.  

Optical observations of OB main sequence stars reported recently that
some of them possess $\sim 1$~kG surface magnetic fields
\citep[e.g.,][]{wad12}, which correspond to the magnetar-class magnetic
flux, $\sim 10^{26}$--$10^{27}$~G cm$^{2}$. A population synthesis
calculation assuming the conservation of 
magnetic flux during the post-main-sequence evolution performed
by \citet{fer06} indicates that $\sim 10$\% of OB stars have
the magnetar-class magnetic flux, while the majority has $\sim$1--2
orders of magnitude weaker ones. 

Stellar evolution calculations so far indicate that the pre-collapse
rotation rate is sensitive to that at the zero-age main sequence
(ZAMS). Whereas \citet{heg05} found that a 15~$M_\odot$ star with the
solar metallicity and the surface rotational velocity of 
$v_{\textrm{rot,ZAMS}}=$200~km~s$^{-1}$ at ZAMS results in the
pre-collapse rotation that may be translated to a pulsar rotation
period of $P_{\textrm{NS}}=11$~ms, a 16~$M_\odot$
star with the solar metallicity and
$v_{\textrm{rot,ZAMS}}=$360~km~s$^{-1}$ computed by 
\citet{woo06} ends up with a pre-collapse rotation corresponding to
$P_{\textrm{NS}}=2.3$--9.7~ms, depending on the unknown mass-loss rate
in the Wolf-Rayet stage.  Meanwhile, \citet{ram13} found that 20~\% of
216 O-type stars in 30 Dor have the surface rotational velocities
larger than 300~km~s$^{-1}$. Combined with 
the results of \citet{woo06}, this implies that rapidly rotating
progenitors that may produce a pulsar rotation period of
milliseconds, may not be so rare even for the solar metallicity.

The influences of magnetic field and rotation on the dynamics of CCSNe
have been so far studied mostly in the regime in which magnetic field
is large and rotation is rapid simultaneously, i.e., the combination
of the magnetar-class magnetic flux and the rotation that would
produce millisecond proto-neutron star (MPNS) has been
assumed. Numerical simulations have shown that 
the magnetic field, which is already quite large initially and is
later amplified by differential rotation, can drive explosions with
the canonical energy of $10^{51}$~erg
\citep[e.g.,][]{yam04,obe06,shi06,saw13a}.   

There are a small number of simulations assuming initially
sub-magnetar-class magnetic flux.
In their axisymmetric (2D) simulations \citet{bur07} and \citet{tak09}
found that even with sub-magnetar-class fields,
$B_{\textrm{pre}}\lesssim 10^{11}$~G at pre-collapse, rapid rotations
corresponding to MPNS can amplify the magnetic fields to produce
well-collimated magneto-driven jets. On the other hand, similar
simulations by \citet{moi06} obtained more or less spherical
explosions. In both cases 
dipole-like magnetic fields assumed at pre-collapse generate explosion
energy smaller than the canonical value. Note, however, that all  
these simulations do not have enough spacial resolutions to capture
magnetorotational instability (MRI), which is expected to occur in
CCSNe and, if true, would drastically amplify the magnetic field in
the timescale of rotation \citep{bal91,aki03}.  

Resolving MRI in CCSNe with sub-magnetar-class magnetic flux demands
quite a fine size of numerical grid with a width of, say, a few 10 m
for $B_{\textrm{pre}}\sim 10^{11}$~G, which should be compared with
the size of iron cores, $\sim 
1000$~km. To deal with high computational cost, some previous
simulations were done in local boxes placed in PNSs
\citep{obe09,mas12}. \citet{saw13b}, on the other hand,
performed global simulations in axisymmetry, deploying as many as
$9300\times 6400$ mesh points and demonstrated
that MRI produces strong magnetic fields on large scales even under the
dynamical background. They also showed that
the magnetic force becomes dynamically important in some
locations. Since their simulations were terminated around 70~ms after
bounce, the consequence for the dynamics thereafter were not clear.
Note that neutrino transport was not included in their simulations,
which is main reason they stopped computations.

In this letter, we investigate the aftermath. In so doing, we assume
again the initial magnetic fields with sub-magnetar-class magnetic
flux and carry out long-term MHD simulations in
axisymmetry with neutrino cooling and heating, albeit simplified,
being taken into account.

In the following, the numerical methods and models are described in
\S\ref{sec.model}, results are presented in \S\ref{sec.result}, and
discussion and conclusion are given in \S\ref{sec.conc}.

\section{Numerical Methods and Models}\label{sec.model}
In the current simulations the following ideal
MHD equations and the equation for the electron number density are
numerically solved by a time-explicit 
Eulerian MHD code, \textit{Yamazakura} \citep{saw13a}:
{\allowdisplaybreaks
\begin{eqnarray}
&&\frac{\partial\rho}{\partial
  t}+\nabla\cdot(\rho\mbox{\boldmath$v$})=0\label{eq.mhd.mass},
\\ 
&&\frac{\partial}{\partial t} (\rho\mbox{\boldmath$v$})+
\nabla\cdot\left(\rho\mbox{\boldmath$v$}\mbox{\boldmath$v$}-
\frac{\mbox{\boldmath$B$}\mbox{\boldmath$B$}}{4\pi}\right)\nonumber\\
&&\hspace{1pc}=-\nabla\left(p+\frac{B^2}{8\pi}\right)-\rho\nabla\Phi 
\label{eq.mhd.mom},
\\  
&&\frac{\partial}{\partial t}\left(e+\frac{\rho
    v^2}{2}+\frac{B^2}{8\pi}\right)
\nonumber\\
&&\hspace{1pc}+\nabla\cdot
\left[
\left(e+p+\frac{\rho
      v^2}{2}+\frac{B^2}{4\pi}\right) 
\mbox{\boldmath$v$}
-\frac{(\mbox{\boldmath$v$}\cdot\mbox{\boldmath$B$}) 
\mbox{\boldmath$B$}}{4\pi}
\right]
\nonumber\\
&&\hspace{4pc}
=-\rho(\nabla\Phi)\cdot\mbox{\boldmath$v$}
+Q_E^{\textrm{abs}}+Q_E^{\textrm{em}}\label{eq.mhd.eng},
\\
&&\frac{\partial\mbox{\boldmath $B$}}{\partial t}=
\nabla\times\left(\mbox{\boldmath$v$}\times\mbox{\boldmath$B$}\right)
\label{eq.mhd.far},
\\
&&\frac{\partial n_e}{\partial
  t}+\nabla\cdot(n_e\mbox{\boldmath$v$})=Q_N^{\textrm{abs}}+Q_N^{\textrm{em}}
\label{eq.ne},
\end{eqnarray}}where $Q_E^{\textrm{abs}}$ and $Q_E^{\textrm{em}}$ are
the rates of energy density change due to $\nu_e$/$\bar{\nu_e}$ absorptions
and emissions, respectively; $Q_E^{\textrm{abs}}$ and
$Q_E^{\textrm{em}}$ are analogues for the change of electron
number density. Following \citet{mur09}, these rates are calculated
based on \citet{jan01}, where we assume a constant
$\nu_e$/$\bar{\nu_e}$ luminosity of $10^{52}$erg s$^{-1}$. 
The other symbols in Equations (\ref{eq.mhd.mass})--(\ref{eq.ne}) have
their usual meanings. The gravitational 
potential $\Phi$ is approximated by the Newtonian
mono-pole gravity. The tabulated nuclear equation of state (EOS)
produced by \citet{she98a, she98b} is adopted in this study.
The electron fraction, $Y_e=n_e m_u/\rho$, where $m_u=1.66\times
10^{-24}$~g is the atomic mass unit, is given by the prescription
suggested by \citet{lie05} until bounce. After that,
Equation~(\ref{eq.ne}) is solved, since such a prescription is no
longer valid. We employ the polar coordinates in two dimensions,
assuming axisymmetry as well as equatorial symmetry.

The collapse is followed for a $15 M_\odot$ progenitor star provided
by S. E. Woosley (1995, private communication). 
A dipole-like magnetic field configuration, which is the same as that
employed by \citet{saw13a}, is initially assumed. Three different
initial strengths 
of the magnetic field are studied, in which the maximum values at
pre-collapse are
$B_{\textrm{pre,max}}=5\times 10^{10}$, $1\times 10^{11}$ and $2\times 10^{11}$~G.
The core is assumed to be rapidly rotating
with the pre-collapse angular velocity profile of 
\begin{eqnarray}
\Omega(r)=\Omega_{0}\frac{r_0^2}{r_0^2+r^2},
\end{eqnarray}
where $r_0=1000$~km and $\Omega_{0}=2.7$~rad s$^{-1}$, which would
produce a MPNS after collapse. The initial
rotational energy divided by the gravitational binding energy,
$T/|W|$, is 0.3~\%. 

As in \citet{saw13b}, we conduct two different sorts of numerical runs,
namely, background (BG) runs and MRI runs. BG runs follow the dynamics
of the central region of the progenitor that covers the entire iron
core and extends upto the radius of
4000~km with $N_r\times N_\theta = 720\times 60$ numerical grids,
which correspond to the radial spatial resolutions of 0.4--23~km.
MRI runs are 
performed with much higher spacial resolutions to capture details of
MRI on small scales. The numerical domain is limited to
$50\leq(r/$km$)\leq500$, and three different spacial resolutions, in
which the innermost grid size, $\Delta_{\textrm{in}}$ (and the numbers
of points, $N_r\times N_\theta$), are 25~m 
($4650\times 3200$), 50~m ($2300\times 1600$), and 100~m ($1160\times
800$) are adopted. Hereafter, we refer to these MRI runs
as H-MRI, M-MRI, L-MRI runs, respectively. The grid
spacing is determined so that the radial and angular grid sizes should
be the same, viz. $\Delta r=r\Delta\theta$, at the innermost and
outermost cells. The snapshots at 6~ms after bounce in BG runs are
utilized as the initial conditions for the corresponding MRI runs.
The results of BG runs are also employed to set the inner and outer
boundary conditions for MRI runs except for $B_r$, 
which is determined so that the divergence-free condition of the
magnetic field at the inner boundary should be satisfied. 

In this letter, we focus on the results for
$B_{\textrm{pre,max}}=5\times 10^{10}$~G, the weakest magnetic
field among our choice, since we are interested in the weak field
regime. Although this is still not a small value, even weaker fields
are not affordable at present because of too-high numerical cost. The
results of the other simulations and their analyses will be 
presented in a forthcoming paper (Sawai \& Yamada 2014, in
preparation). The details of numerical technics will also be
described there.

\section{Results}\label{sec.result}
\begin{figure}
\epsscale{1}
\plotone{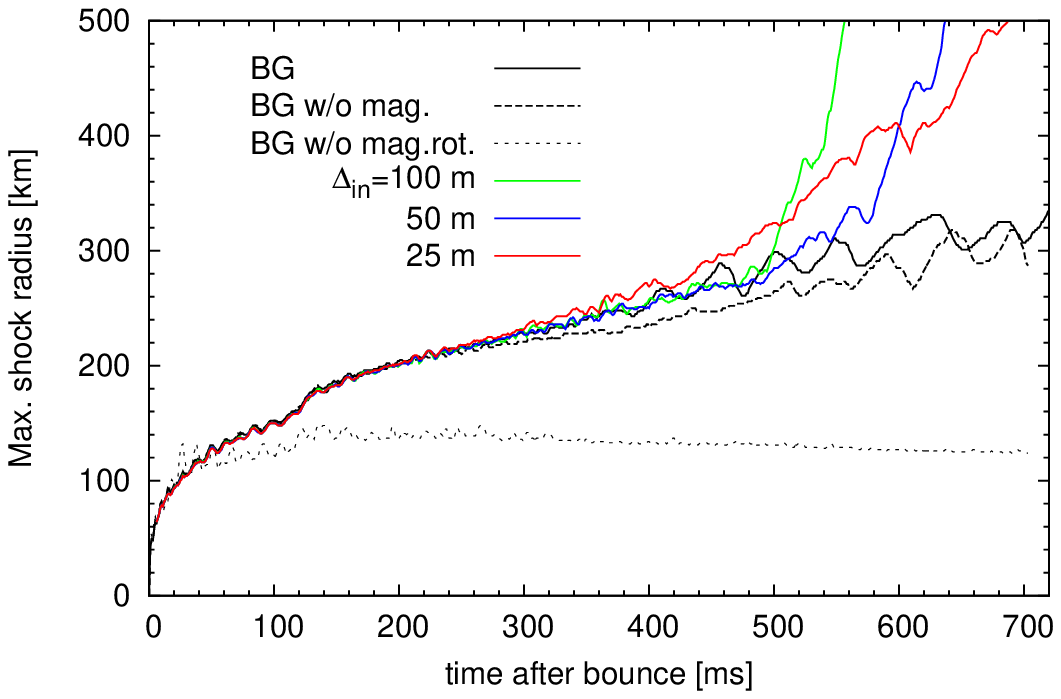}
\plotone{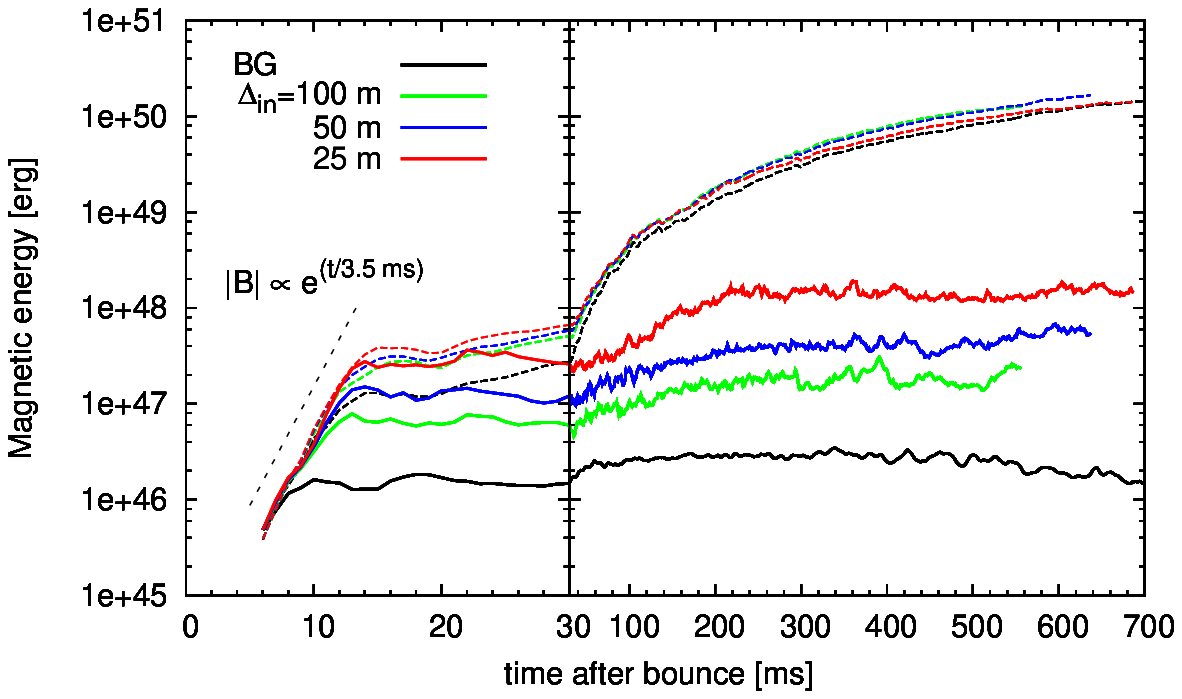}
\caption{Top panel: Evolutions of the maximum shock radii. Bottom panel:
  Evolutions of the magnetic energies integrated over
  $50\leq(r/$km$)\leq500$ for the poloidal component (solid lines) and
toroidal component.} 
\label{fig.rsh}
\end{figure}

The postbounce evolution of the BG run with
$B_{\textrm{pre,max}}=5\times 10^{10}$~G shows a gradual increase of
the maximum shock radius, which reaches $\sim 300$~km at 700~ms after
bounce (compare the black-solid line in the top panel of
Figure~\ref{fig.rsh}). We performed two additional
simulations for comparison, namely, one without both magnetic field
and rotation and the other without magnetic field alone. The shock
stalls around 100~km in the former case,  
while the latter case shows an evolution of the maximum shock radius
that is not much different from the one in the BG run except for a
little slower pace of increase (see the black-dotted and -dashed lines
in the top panel of Figure~\ref{fig.rsh}). 
The very small difference between the simulations with and without magnetic
field implies that the influence of magnetic field is insignificant in
the BG run. Although a substantial fraction of the collapsed core is
found to be unstable to MRI in the BG run, MRI is not resolved in
this simulation due to insufficient numerical resolution. In fact wave
length of the mode with the greatest growth rate is only two-grid wide.

The bottom panel of Figure~\ref{fig.rsh} shows that
the poloidal magnetic fields in the MRI runs grow exponentially around
10~ms after bounce, an indication that MRI is in operation. In fact
the growth timescale of $\sim 3.5$~ms in H-MRI run is roughly consistent
with the one expected from the linear analysis. It is noted that the
fastest growing mode is mostly resolved by more than 10 grid
points. The exponential growth is saturated at values that are much
greater than those before the amplification: the poloidal magnetic
field strength around the radius of 60~km in the vicinity of the pole
increases from $\sim 10^{13}$~G to $\sim 10^{14}$~G during this period
of time. Thereafter the poloidal fields remain almost constant whereas
the toroidal fields are gradually amplified by winding. As found in
\citet{saw13b}, the saturation level of the poloidal component is
larger for higher resolutions possibly due to smaller numerical
diffusivity. This is also the case for the toroidal component just
after the end of linear growth phase. It is interesting, however, the
toroidal field strengths become more or less similar at later times
irrespective of numerical resolutions. It is unfortunately evident
that even higher resolutions are needed to see a convergence in the
sub-dominant poloidal field.

\begin{figure}
\epsscale{1}
\plotone{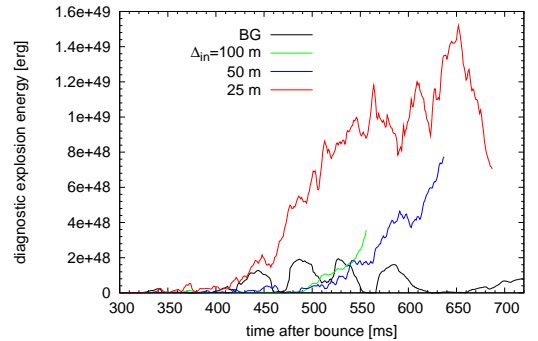}
\caption{Evolutions of the diagnostic explosion energies defined in
  the text.}
\label{fig.eexp}
\end{figure}

As can be seen from the top panel of Figure~\ref{fig.rsh}, the MRI
runs obtain more rapid increases of the maximum shock
radius compared with the BG run. The evolutions of the
diagnostic explosion energy, which is the energy integrated over the
fluid elements that move outwards with positive energies, also
indicate that the shock expansion leads to explosions in the MRI runs
but not in the BG run (see Fig~\ref{fig.eexp}).
While the shock radius grows fastest in the L-MRI run,
the H-MRI run is likely to give the strongest explosion. As discussed
bellow, we found that this is a consequence of efficient angular
momentum transfer by larger saturation fields obtained in higher
resolution runs.

\begin{figure*}
\epsscale{1.1}
\plottwo{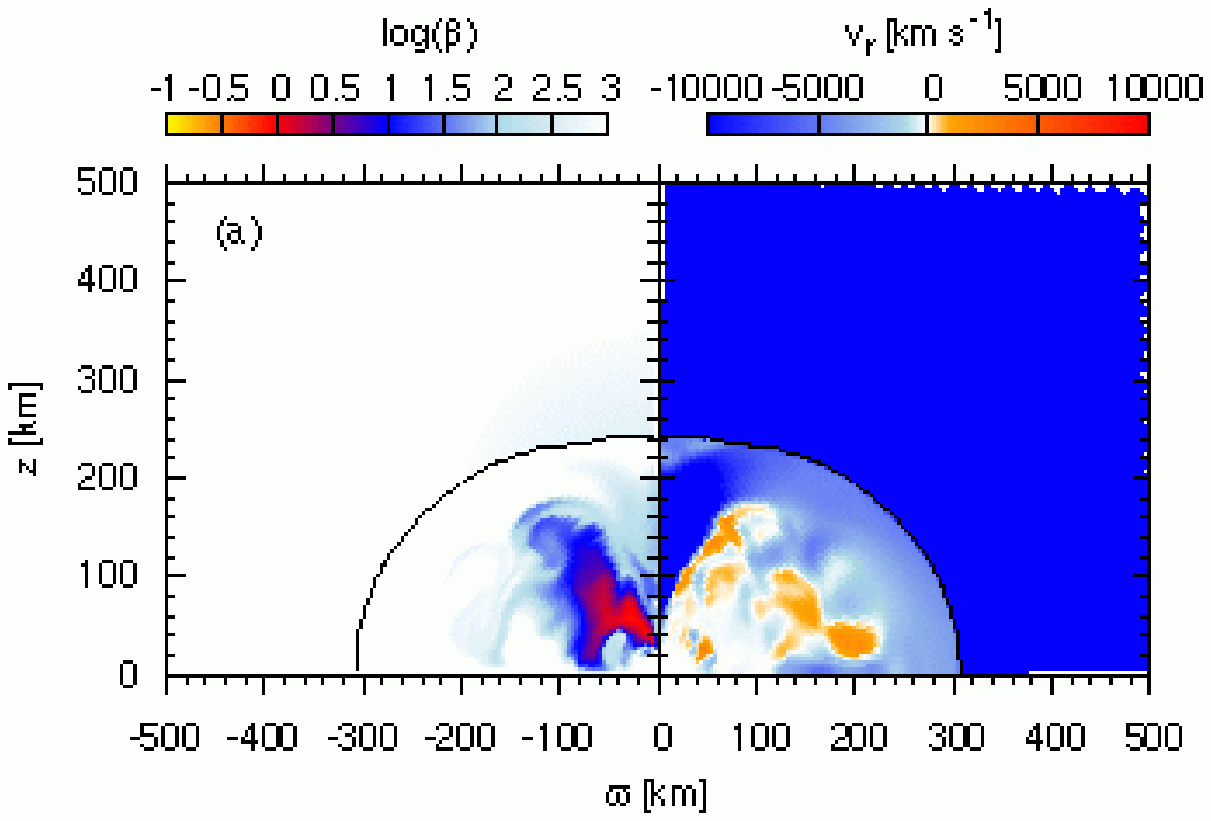}{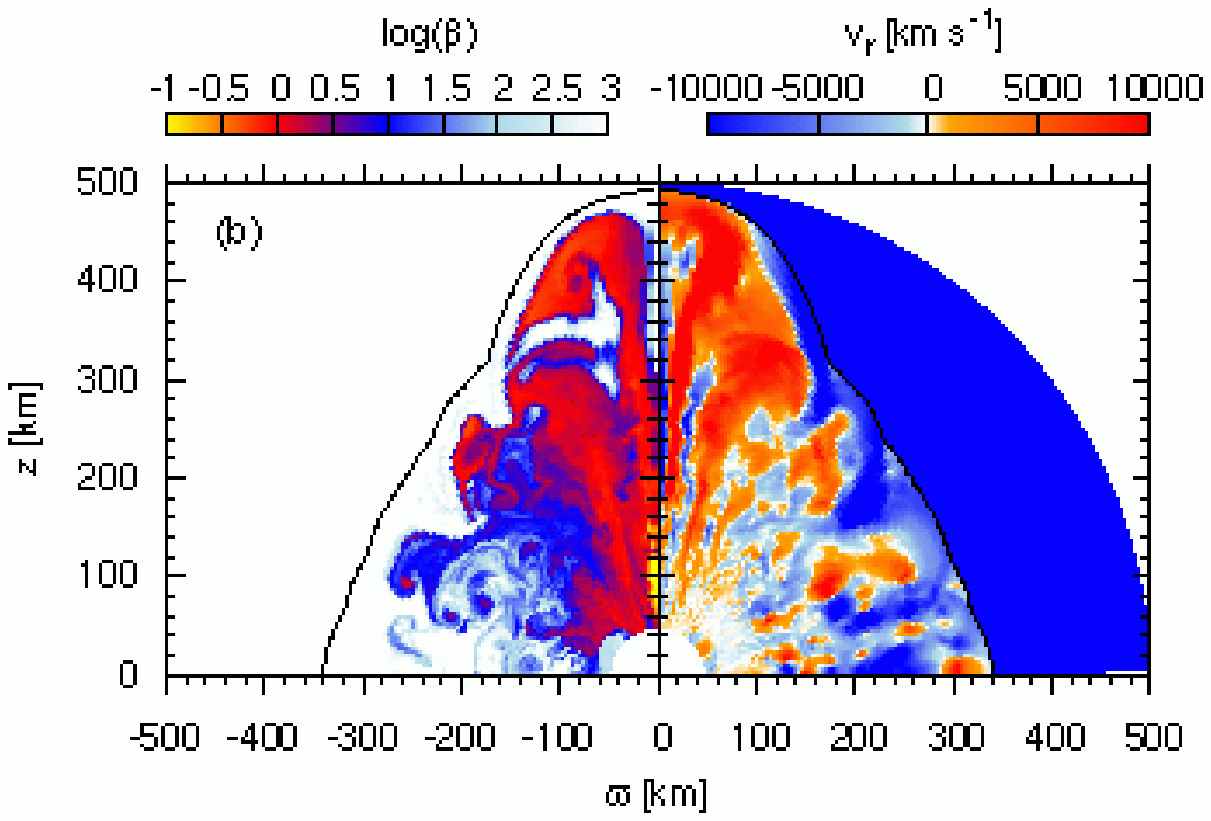}
\plottwo{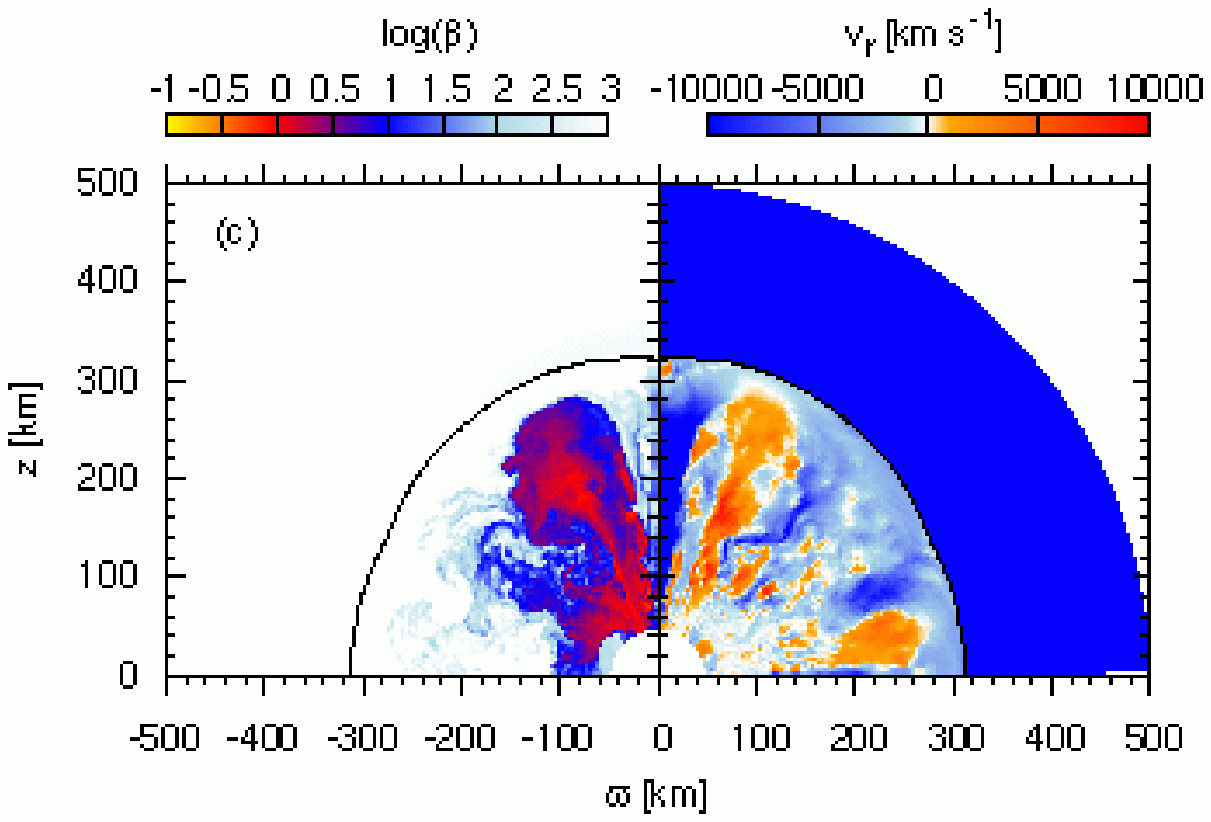}{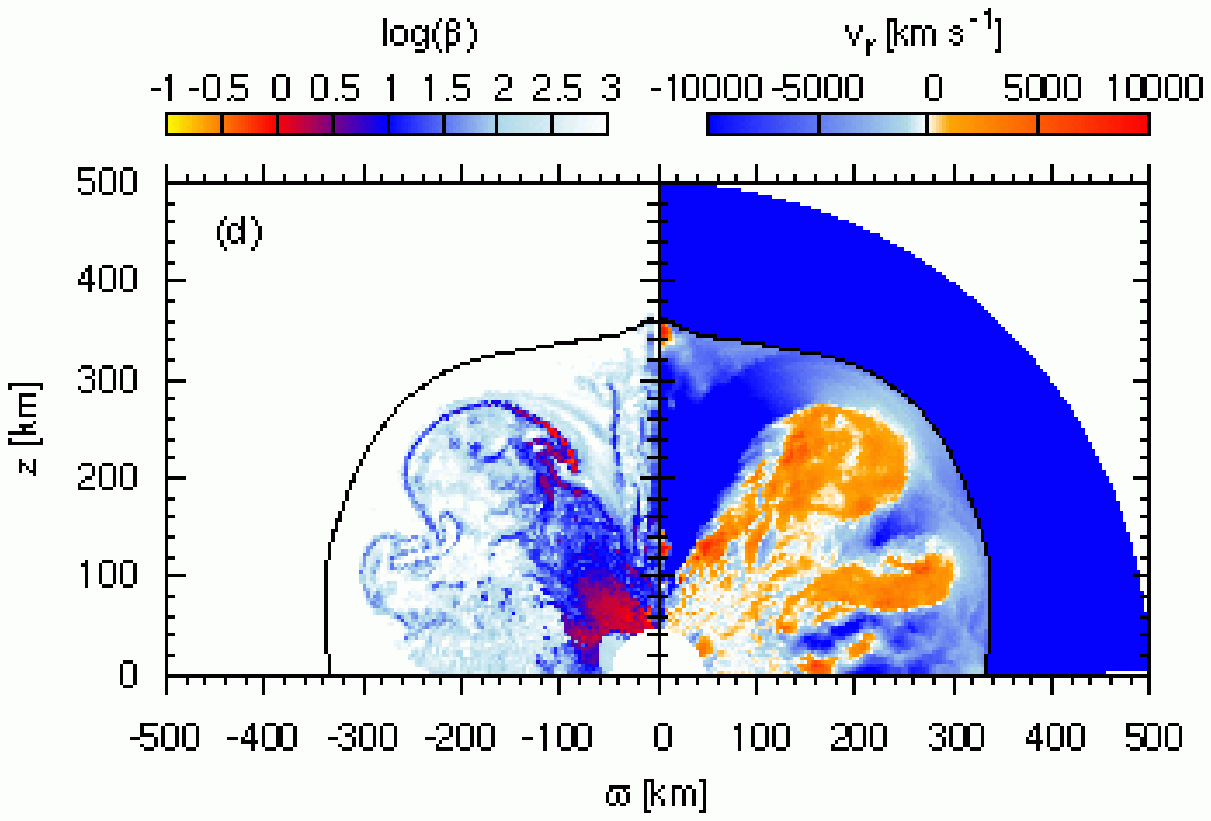}
\caption{Color maps of plasma beta (the left half) and
  radial velocity magnitude (the right half), and the contour of the
  shock surfaces (black line) at 555 ms after bounce for BG run (a),
  L-MRI run (b), M-MRI run (c), and H-MRI run (d).}
\label{fig.betavrad}
\vspace{1pc}
\end{figure*}

Figure~\ref{fig.betavrad} displays the color maps of plasma beta,
$\beta=8\pi p/B^2$, and radial velocity for all simulations at
555 ms after bounce. One can recognize a qualitative change in
dynamics between the BG run and the L-MRI run: a low-$\beta$ region
around the pole extend further and mass ejection instead of accretion
occurs in the L-MRI run. Interestingly, however, low-$\beta$ region shrinks
again with the further improvement of resolution. As a result the jet
disappears and mass accretion occurs again around the pole in the
M-MRI and H-MRI runs.
On the other hand, mass ejection around mid-latitude becomes more
prominent in H-MRI run (compare panel (b) and
(d) of Figure~\ref{fig.betavrad}). Although this mid-latitude ejection
is slower than the polar 
ejection observed in the L-MRI run, the former results in a larger
explosion energy as seen in Figure~\ref{fig.eexp}, since the amount of
ejected mass is larger. Note that panel (d) indicates that the
mid-latitude ejection is not powered by magnetic pressure, since its
plasma beta is $\sim 10$--1000.

\begin{figure*}
\epsscale{1}
\plottwo{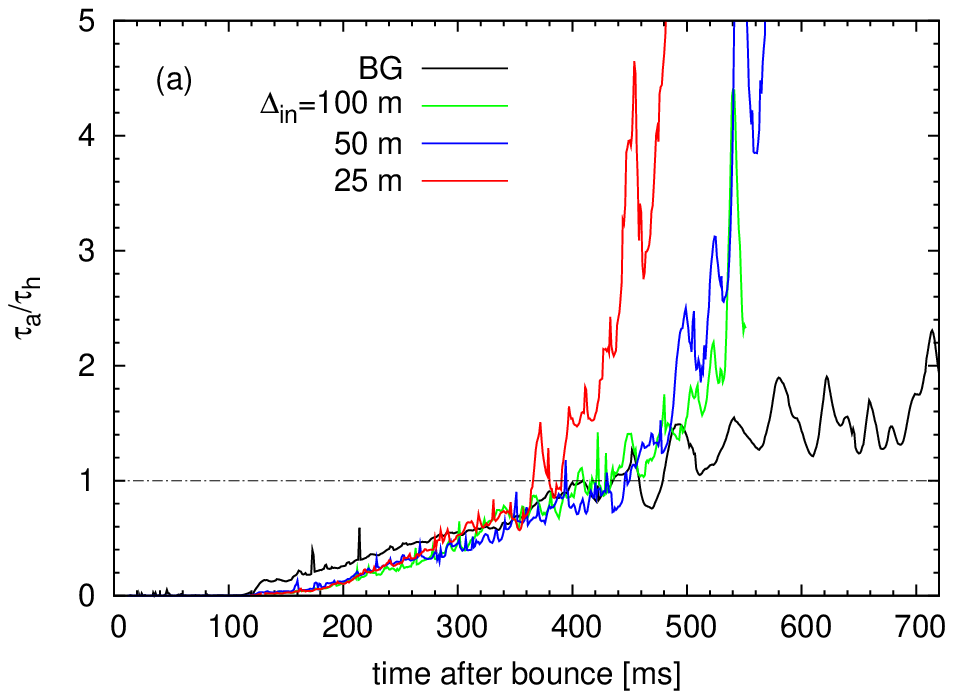}{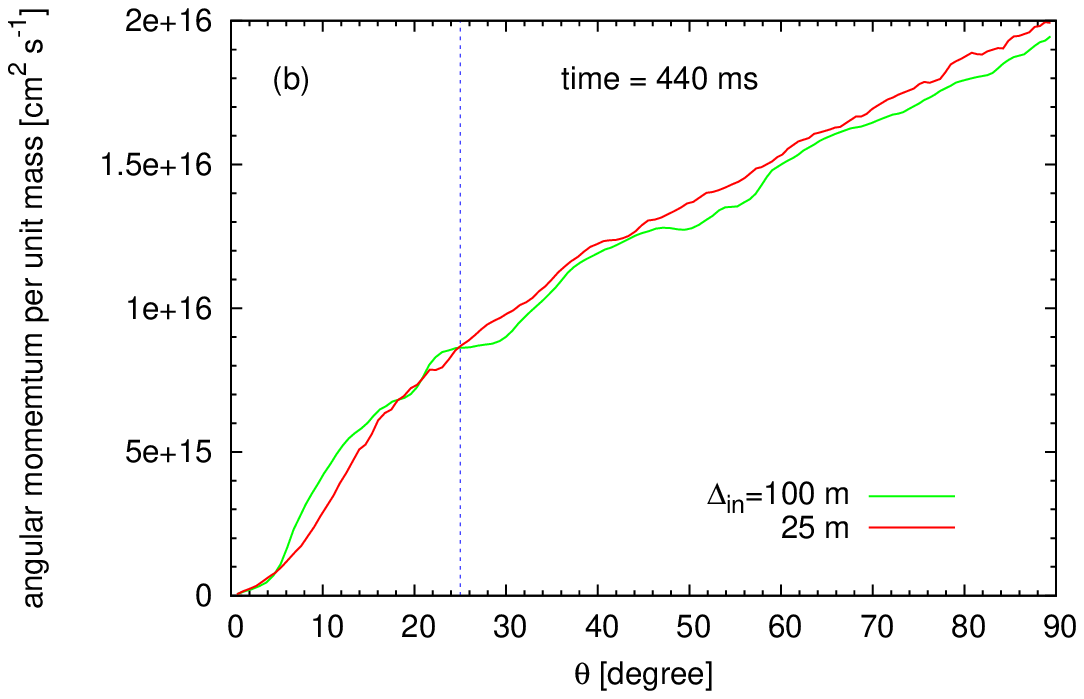}
\epsscale{1.1}
\plottwo{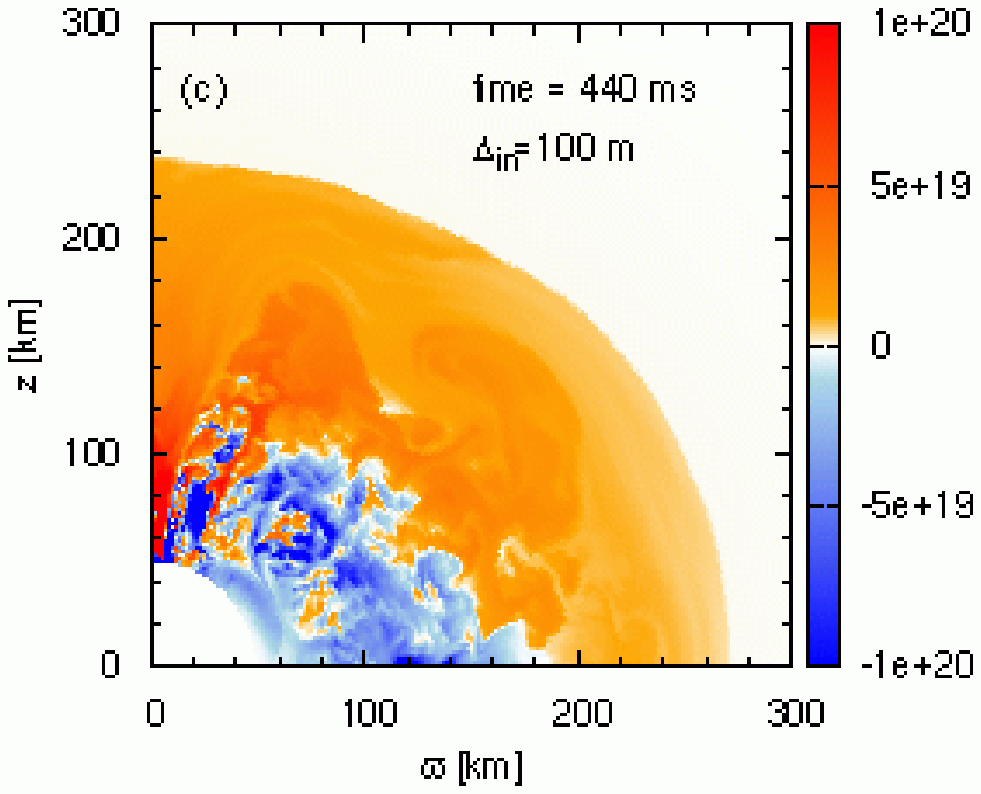}{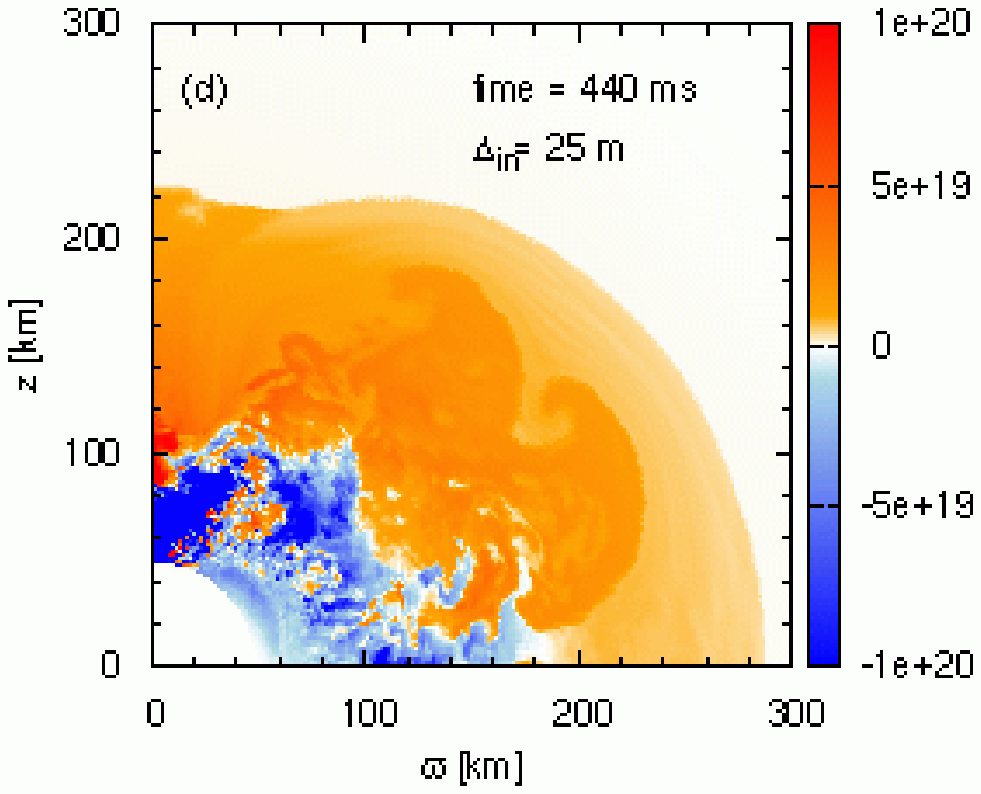}
\caption{(a):Evolutions of the ratio of the advection timescale to
  heating timescale, whose definitions are given in the text. (b):
  Angular distributions of the radially averaged angular momentum per
  unit mass at 440~ms after bounce for L-MRI run (green line) and
  H-MRI run (red line). Bottom panels: Color maps of the net heating
  rate per unit mass at 440~ms after bounce for L-MRI run (c) and
  H-MRI run (d).}
\label{fig.heat}
\vspace{2pc}
\end{figure*}

The comparison between the advection timescale, $\tau_{\textrm{a}}$,
during which matter traverses the gain region, and the heating
timescale, $\tau_{\textrm{h}}$, within which matter gains enough
energy to overcome gravity, is one of the rough measure to judge
whether the neutrino heating plays an important role in driving
explosion \citep{tho00}. Following \citet{dol13}, we define the
advection timescale as  
\begin{equation}
\tau_{\textrm{a}}=\int^{R_{\textrm{gain}}}_{R_{\textrm{sh}}}
\frac{dr}{\langle\langle v_r\rangle\rangle},
\end{equation}
where the double angle bracket denotes that the solid-angle average
and time average over the interval of 10~ms are taken. 
$R_{\textrm{sh}}$ is taken as the mean shock radius, whereas
$R_{\textrm{gain}}$ is defined as the innermost radius at
which the solid-angle averaged net heating is positive. The heating
timescale is defined as
\begin{equation}
\tau_{\textrm{h}}=
\frac{4\pi\int^{R_{\textrm{gain}}}_{R_{\textrm{sh}}}
\langle e+\frac{\rho v^2}{2}+\frac{B^2}{8\pi}+\rho\Phi\rangle r^2 dr}
{4\pi\int^{R_{\textrm{gain}}}_{R_{\textrm{sh}}}
\langle Q_E^{\textrm{em}}+ Q_E^{\textrm{abs}}\rangle r^2dr},
\end{equation}
where the angle brackets denote that the solid-angle average is taken.
Panel (a) of Figure~\ref{fig.heat} displays the temporal variations of
$\tau_{\textrm{a}}/\tau_{\textrm{h}}$. It is shown that the neutrino
heating is important in all numerical runs and that the heating
efficiency is higher in the MRI runs than in the BG run. While the
L-MRI and M-MRI runs have similar ratios of
$\tau_{\textrm{a}}/\tau_{\textrm{h}}$, the H-MRI run has much larger
ratio after $\sim 400$~ms, suggesting that the
mid-latitude mass ejection observed in the H-MRI run (panel (d) of
Figure~\ref{fig.betavrad}) is driven by neutrino heating.

We think that the high $\tau_{\textrm{a}}/\tau_{\textrm{h}}$ in the
H-MRI run is caused by an efficient angular momentum transfer. In
fact, panel (b) of Figure~\ref{fig.heat} shows that the angular
momentum per unit mass in the H-MRI run is larger than that in the
L-MRI run for $\theta\gtrsim 25^\circ$, and vice versa for
$\theta\lesssim 25^\circ$, which is probably a consequence of more
efficient angular momentum transfer in H-MRI run. The 
rate of angular momentum transfer  
is proportional to the product of the poloidal and toroidal components
of magnetic field. Although the saturation level of the  toroidal
component is similar among all the simulations
presented in this letter, the larger poloidal field at saturation
in the H-MRI run than in the other runs (bottom panel of
Figure~\ref{fig.rsh}) makes angular momentum transfer more efficient.
This then leads to the widening of the heating region at low latitudes
in the same model due to the extra-rotational support (compare the
bottom panels of Figure~\ref{fig.heat}.) As a consequence, the
advection timescale becomes long, and thus the heating efficiency gets
higher. Although the heating region is thicker around the pole in the
L-MRI run than in the H-MRI run, this 
contributes little to the volume integrated heating rate because of the
small mass in the region.

The fact that the mass ejection around the pole in the L-MRI run turns
to the mass accretion in higher resolution run (see Figure~\ref{fig.betavrad})
could be also accounted for by the efficiency of angular momentum
transfer. Since the rotational support around the pole is weaker for
the higher resolution runs, matter accretes more easily. This argument
seems to contradict the fact that the mass accretion in the BG run
turns to the mass ejection in the L-MRI run in the first place. The
point here is that there is a trade-off between the gain in the
magnetic stress and the loss in the centrifugal support.

\section{Discussion and Conclusion}\label{sec.conc}
We conducted two dimensional high-resolution global MHD simulations of
CCSNe, assuming that a massive star core is rather weakly magnetized
and rapidly rotating prior to collapse. Taking neutrino heating and
cooling into account approximately, we followed the evolutions long
after core bounce and studied the dynamical consequences of MRI. We
found that the magnetic field amplified by MRI indeed affects the
dynamics thereafter substantially enhances the explosion. In fact, it
efficiently transfers angular momentum from high to low latitudes
and expands the heating region at low latitudes not by magnetic
stress but by centrifugal force. This then enhances the heating
efficiency and leads to explosion. This is a new explosion mechanism,
which we expect to work generally in weakly magnetized, rapidly
rotating stellar cores. In fact, although even our highest-resolution
run has not yet achieved numerical convergence particularly in the
sub-dominant poloidal component, further improvement of resolution
would result in more efficient angular momentum transfer and thus in
higher neutrino heating, which would make shock revival by neutrino
heating even easier.

Previous simulations of weakly magnetized, rapidly-rotating cores by
\citet{bur07} and \citet{tak09} found magneto-driven jets formed along
the rotation axis, which are analogous to what we found in
the L-MRI run. Since the 
spacial resolution in \citet{bur07} is in between our BG and L-MRI
runs, where the transition from mass accretion to ejection occurs near
the pole, their results do not contradict ours. What they observed may
be an artifact by insufficient numerical resolutions as demonstrated
in the current study. Although \citet{bur07} 
claimed that the neutrino heating is subdominant
(contributing 10--25~\% to the explosion energy), our simulations
indicate that it is actually predominant. The spacial resolution in
\citet{tak09} is coarser than our BG run. The
reason for the jet formations in their simulations may be because
they assumed strong differential rotation prior to collapse with a
typical scale height of 100~km in the $\varpi$ direction. 

The diagnostic explosion energy obtained in our highest
resolution simulation (H-MRI run) is $\sim 10^{49}$~erg, which is far
smaller than the canonical value, $10^{51}$~erg. It should be noted
that the shock front reaches only the radius of 500~km at the end of
the simulation. It is way too early to estimate the final value of the
explosion energy, since it may increase as the shock front propagates
outward. Further improvement of the spacial resolution may also
increase the explosion energy, since more 
efficient angular momentum transfer is expected. Moreover, the
neutrino luminosity of $10^{52}$erg s$^{-1}$ assumed in this
study may be a bit too low. In fact, a recent simulation by
\citet{bru13} obtained $\nu_e$/$ \bar{\nu_e}$ of several $10^{52}$erg
s$^{-1}$ for the first 200~ms after bounce, which decreases to
$\approx 10^{52}$erg s$^{-1}$ only later at $\sim 700$~ms after
bounce. Larger neutrino luminosities will certainly result in larger
explosion energies. Apart from the 
quantitative estimation of the explosion energy, what is most important 
here is the finding that even a weak magnetic field added to a rapidly
rotating core may make the neutrino-driven explosion easier.

\acknowledgments
H.S. is grateful to Shun Furusawa, Ryosuke Hirai, Wakana Iwakami,
Hiroki Nagakura, Ko 
Nakamura, and Yu Yamamoto for fruitful discussion and useful advice.
The numerical computations were carried out on XC30 at CfCA, National
Astronomical Observatory of Japan, and on SR16000 at YITP, Kyoto University.
This work is supported by a Grant-in-Aid for Scientific Research from
the Ministry of Education, Culture, Sports, Science and Technology,
Japan (21840050, 24103006, 24244036.)

\end{document}